\documentclass[preprint,aps]{revtex4}
\usepackage{dcolumn}
\usepackage{bm}
\usepackage{url}
\usepackage{color}
\usepackage[T1]{fontenc}
\usepackage{lineno}
\usepackage{epsfig}
\newcommand{\isotope}[2]{\ensuremath{{^{#1}}\textrm{#2}}}
\newcommand{\reaction}[4]{\ensuremath{ {#1} ( {#2} , {#3} ) {#4} } }

\newcommand{\geant}{\ensuremath{\textsc{Geant4}}}

\newcommand{\val}[2]{\ensuremath{#1(#2)}}

\newcommand{\tref}{Ref.~}
\newcommand{\tfig}{Fig.~}
\newcommand{\ttab}{Table~}
\newcommand{\tsec}{Sec.~}

\begin{document}

\preprint{APS/123-QED}

\title{High-spin structure in  $^{40}$K}

\author{P.-A.~S{\"o}derstr{\"o}m}\affiliation{Department of Physics and Astronomy, Uppsala University, SE-75120 Uppsala, Sweden}\affiliation{RIKEN Nishina Center, 2-1 Hirosawa, Wako-shi, Saitama 351-0198, Japan}
\author{F.~Recchia}\affiliation{INFN Sezione di Padova, I-35131 Padova, Italy}\affiliation{Dipartimento di Fisica, Universit\`{a} di Padova, I-35131 Padova, Italy}
\author{J.~Nyberg}\affiliation{Department of Physics and Astronomy, Uppsala University, SE-75120 Uppsala, Sweden}
\author{A.~Gadea}\affiliation{Instituto de F\'{\i}sica Corpuscular, CSIC-Universidad de Valencia, E-46071 Valencia, Spain}
\author{S.M.~Lenzi}\affiliation{INFN Sezione di Padova, I-35131 Padova, Italy}\affiliation{Dipartimento di Fisica, Universit\`{a} di Padova, I-35131 Padova, Italy}
\author{A.~Poves}\affiliation{Departamento de F\'{\i}sica Te\'{o}rica e IFT-UAM/CSIC, Universidad Aut\'{o}noma de Madrid, E-28049 Madrid, Spain}
\author{A.~Ata\c{c}}\affiliation{Department of Physics, Faculty of Science, Ankara University, 06100 Tando\u{g}an, Ankara, Turkey}
\author{S.~Aydin}\affiliation{INFN Sezione di Padova, I-35131 Padova, Italy}\affiliation{Dipartimento di Fisica, Universit\`{a} di Padova, I-35131 Padova, Italy}\affiliation{Department of Physics, Faculty of Science and Art, Aksaray University, Aksaray 68100, Turkey}
\author{D.~Bazzacco}\affiliation{INFN Sezione di Padova, I-35131 Padova, Italy}
\author{P.~Bednarczyk}\affiliation{The Henryk Niewodnicza\'{n}ski Institute of Nuclear Physics, Polish Academy of Sciences, ul. Radzikowskiego 152, 31-342 Krak\'{o}w, Poland}
\author{M.~Bellato}\affiliation{INFN Sezione di Padova, I-35131 Padova, Italy}
\author{B.~Birkenbach}\affiliation{Institut f\"{u}r Kernphysik, Universit\"{a}t zu K\"{o}ln, Z\"{u}lpicher Str. 77, D-50937 K\"{o}ln, Germany}
\author{D.~Bortolato}\affiliation{Istituto Nazionale di Fisica Nucleare, Laboratori Nazionali di Legnaro, I-35020 Legnaro, Italy}
\author{A.J.~Boston}\affiliation{Oliver Lodge Laboratory, The University of Liverpool, Liverpool, L69 7ZE, UK}
\author{H.C.~Boston}\affiliation{Oliver Lodge Laboratory, The University of Liverpool, Liverpool, L69 7ZE, UK}
\author{B.~Bruyneel}\affiliation{Institut f\"{u}r Kernphysik, Universit\"{a}t zu K\"{o}ln, Z\"{u}lpicher Str. 77, D-50937 K\"{o}ln, Germany}
\author{D.~Bucurescu}\affiliation{Horia Hulubei National Institute of Physics and Nuclear Engineering  P.O.BOX MG-6, Bucharest - Magurele, Romania}
\author{E.~Calore}\affiliation{Istituto Nazionale di Fisica Nucleare, Laboratori Nazionali di Legnaro, I-35020 Legnaro, Italy}
\author{B.~Cederwall}\affiliation{Department of Physics, Royal Institute of Technology, SE-10691 Stockholm, Sweden}
\author{L.~Charles}\affiliation{Universit\'{e}~de~Strasbourg,~IPHC,~23~rue~du~Loess,~67037~Strasbourg,~France CNRS,~UMR~7178,~67037~Strasbourg,~France}
\author{J.~Chavas}\affiliation{CEA Saclay, IRFU/Service de Physique Nucl\'{e}aire, F-91191 Gif-sur-Yvette Cedex, France}
\author{S.~Colosimo}\affiliation{Oliver Lodge Laboratory, The University of Liverpool, Liverpool, L69 7ZE, UK}
\author{F.C.L.~Crespi}\affiliation{INFN Sezione di Milano, I-20133 Milano, Italy}\affiliation{Dipartimento di Fisica, Universit\`{a} di Milano, I-20133 Milano, Italy}
\author{D.M.~Cullen}\affiliation{Nuclear Physics Group, Schuster Laboratory, University of Manchester, M13 9PL, UK }
\author{G.~de~Angelis}\affiliation{Istituto Nazionale di Fisica Nucleare, Laboratori Nazionali di Legnaro, I-35020 Legnaro, Italy}
\author{P.~D\'{e}sesquelles}\affiliation{CSNSM, CNRS/IN2P3 and Universit\'{e} Paris-Sud, F-91405 Orsay Campus, France}
\author{N.~Dosme}\affiliation{CSNSM, CNRS/IN2P3 and Universit\'{e} Paris-Sud, F-91405 Orsay Campus, France}
\author{G.~Duch\^{e}ne}\affiliation{Universit\'{e}~de~Strasbourg,~IPHC,~23~rue~du~Loess,~67037~Strasbourg,~France CNRS,~UMR~7178,~67037~Strasbourg,~France}
\author{J.~Eberth}\affiliation{Institut f\"{u}r Kernphysik, Universit\"{a}t zu K\"{o}ln, Z\"{u}lpicher Str. 77, D-50937 K\"{o}ln, Germany}
\author{E.~Farnea}\affiliation{INFN Sezione di Padova, I-35131 Padova, Italy}
\author{F.~Filmer}\affiliation{Oliver Lodge Laboratory, The University of Liverpool, Liverpool, L69 7ZE, UK}
\author{A.~G\"{o}rgen}\affiliation{CEA Saclay, IRFU/Service de Physique Nucl\'{e}aire, F-91191 Gif-sur-Yvette Cedex, France}\affiliation{Department of Physics, University of Oslo, Oslo, Norway}
\author{A.~Gottardo}\affiliation{Istituto Nazionale di Fisica Nucleare, Laboratori Nazionali di Legnaro, I-35020 Legnaro, Italy}
\author{J.~Gr\k{e}bosz}\affiliation{The Henryk Niewodnicza\'{n}ski Institute of Nuclear Physics, Polish Academy of Sciences, ul. Radzikowskiego 152, 31-342 Krak\'{o}w, Poland}
\author{M.~Gulmini}\affiliation{Istituto Nazionale di Fisica Nucleare, Laboratori Nazionali di Legnaro, I-35020 Legnaro, Italy}
\author{H.~Hess}\affiliation{Institut f\"{u}r Kernphysik, Universit\"{a}t zu K\"{o}ln, Z\"{u}lpicher Str. 77, D-50937 K\"{o}ln, Germany}
\author{T.A.~Hughes}\affiliation{Oliver Lodge Laboratory, The University of Liverpool, Liverpool, L69 7ZE, UK}
\author{G.~Jaworski}\affiliation{Faculty of Physics, Warsaw University of Technology, Koszykowa 75, 00-662 Warszawa, Poland}\affiliation{Heavy Ion Laboratory, University of Warsaw, ul. Pasteura 5A, 02-093 Warszawa, Poland}
\author{J.~Jolie}\affiliation{Institut f\"{u}r Kernphysik, Universit\"{a}t zu K\"{o}ln, Z\"{u}lpicher Str. 77, D-50937 K\"{o}ln, Germany}
\author{P.~Joshi}\affiliation{Department of Physics, University of York, Heslington, York, YO10 5DD, UK}
\author{D.S.~Judson}\affiliation{Oliver Lodge Laboratory, The University of Liverpool, Liverpool, L69 7ZE, UK}
\author{A.~Jungclaus}\affiliation{Instituto de Estructura de la Materia, CSIC, E-28006 Madrid, Spain}
\author{N.~Karkour}\affiliation{CSNSM, CNRS/IN2P3 and Universit\'{e} Paris-Sud, F-91405 Orsay Campus, France}
\author{M.~Karolak}\affiliation{CEA Saclay, IRFU/Service de Physique Nucl\'{e}aire, F-91191 Gif-sur-Yvette Cedex, France}
\author{R.S.~Kempley}\affiliation{Department of Physics, University of Surrey, Guildford, GU2 7XH, UK}
\author{A.~Khaplanov}\affiliation{Department of Physics, Royal Institute of Technology, SE-10691 Stockholm, Sweden}
\author{W.~Korten}\affiliation{CEA Saclay, IRFU/Service de Physique Nucl\'{e}aire, F-91191 Gif-sur-Yvette Cedex, France}
\author{J.~Ljungvall}\affiliation{CSNSM, CNRS/IN2P3 and Universit\'{e} Paris-Sud, F-91405 Orsay Campus, France}\affiliation{CEA Saclay, IRFU/Service de Physique Nucl\'{e}aire, F-91191 Gif-sur-Yvette Cedex, France}
\author{S.~Lunardi}\affiliation{INFN Sezione di Padova, I-35131 Padova, Italy}\affiliation{Dipartimento di Fisica, Universit\`{a} di Padova, I-35131 Padova, Italy}
\author{A.~Maj}\affiliation{The Henryk Niewodnicza\'{n}ski Institute of Nuclear Physics, Polish Academy of Sciences, ul. Radzikowskiego 152, 31-342 Krak\'{o}w, Poland}
\author{G.~Maron}\affiliation{Istituto Nazionale di Fisica Nucleare, Laboratori Nazionali di Legnaro, I-35020 Legnaro, Italy}
\author{W.~M\k{e}czy\'{n}ski}\affiliation{The Henryk Niewodnicza\'{n}ski Institute of Nuclear Physics, Polish Academy of Sciences, ul. Radzikowskiego 152, 31-342 Krak\'{o}w, Poland}
\author{R.~Menegazzo}\affiliation{INFN Sezione di Padova, I-35131 Padova, Italy}
\author{D.~Mengoni}\affiliation{INFN Sezione di Padova, I-35131 Padova, Italy}\affiliation{Dipartimento di Fisica, Universit\`{a} di Padova, I-35131 Padova, Italy}\affiliation{School of Engineering, University of the West of Scotland, High Street, Paisley, PA1 2BE UK}
\author{C.~Michelagnoli}\affiliation{INFN Sezione di Padova, I-35131 Padova, Italy}\affiliation{Dipartimento di Fisica, Universit\`{a} di Padova, I-35131 Padova, Italy}
\author{P.~Molini}\affiliation{INFN Sezione di Padova, I-35131 Padova, Italy}\affiliation{Dipartimento di Fisica, Universit\`{a} di Padova, I-35131 Padova, Italy}
\author{D.R.~Napoli}\affiliation{Istituto Nazionale di Fisica Nucleare, Laboratori Nazionali di Legnaro, I-35020 Legnaro, Italy}
\author{P.J.~Nolan}\affiliation{Oliver Lodge Laboratory, The University of Liverpool, Liverpool, L69 7ZE, UK}
\author{M.~Norman}\affiliation{Oliver Lodge Laboratory, The University of Liverpool, Liverpool, L69 7ZE, UK}
\author{A.~Obertelli}\affiliation{CEA Saclay, IRFU/Service de Physique Nucl\'{e}aire, F-91191 Gif-sur-Yvette Cedex, France}
\author{Zs.~Podolyak}\affiliation{Department of Physics, University of Surrey, Guildford, GU2 7XH, UK}
\author{A.~Pullia}\affiliation{INFN Sezione di Milano, I-20133 Milano, Italy}\affiliation{Dipartimento di Fisica, Universit\`{a} di Milano, I-20133 Milano, Italy}
\author{B.~Quintana}\affiliation{Laboratorio de Radiaciones Ionizantes, Universidad de Salamanca, E-37008 Salamanca, Spain}
\author{N.~Redon}\affiliation{Universit\'{e} de Lyon, Universit\'{e} Lyon-1, CNRS/IN2P3, IPNL, F-69622 Villeurbanne Cedex, France}
\author{P.H.~Regan}\affiliation{Department of Physics, University of Surrey, Guildford, GU2 7XH, UK}
\author{P.~Reiter}\affiliation{Institut f\"{u}r Kernphysik, Universit\"{a}t zu K\"{o}ln, Z\"{u}lpicher Str. 77, D-50937 K\"{o}ln, Germany}
\author{A.P.~Robinson}\affiliation{Nuclear Physics Group, Schuster Laboratory, University of Manchester, M13 9PL, UK }
\author{E.~\c{S}ahin}\affiliation{Istituto Nazionale di Fisica Nucleare, Laboratori Nazionali di Legnaro, I-35020 Legnaro, Italy}
\author{J.~Simpson}\affiliation{STFC Daresbury Laboratory, Daresbury, Warrington, WA44AD, UK}
\author{M.D.~Salsac}\affiliation{CEA Saclay, IRFU/Service de Physique Nucl\'{e}aire, F-91191 Gif-sur-Yvette Cedex, France}
\author{J.F.~Smith}\affiliation{School of Engineering, University of the West of Scotland, High Street, Paisley, PA1 2BE UK}
\author{O.~St\'{e}zowski}\affiliation{Universit\'{e} de Lyon, Universit\'{e} Lyon-1, CNRS/IN2P3, IPNL, F-69622 Villeurbanne Cedex, France}
\author{Ch.~Theisen}\affiliation{CEA Saclay, IRFU/Service de Physique Nucl\'{e}aire, F-91191 Gif-sur-Yvette Cedex, France}
\author{D.~Tonev}\affiliation{Bulgarian Academy of Sciences, Institute for Nuclear Research and Nuclear Energy, 1784, Sofia, Bulgaria}
\author{C.~Unsworth}\affiliation{Oliver Lodge Laboratory, The University of Liverpool, Liverpool, L69 7ZE, UK}
\author{C.A.~Ur}\affiliation{INFN Sezione di Padova, I-35131 Padova, Italy}
\author{J.J.~Valiente-Dob\'{o}n}\affiliation{Istituto Nazionale di Fisica Nucleare, Laboratori Nazionali di Legnaro, I-35020 Legnaro, Italy}
\author{A.~Wiens}\affiliation{Institut f\"{u}r Kernphysik, Universit\"{a}t zu K\"{o}ln, Z\"{u}lpicher Str. 77, D-50937 K\"{o}ln, Germany}

\date{\today}

\begin{abstract}
High-spin states of $^{40}$K have been populated in the fusion-evaporation reaction $^{12}$C($^{30}$Si,np)$^{40}$K and studied by means of $\gamma$-ray spectroscopy techniques using one AGATA triple cluster detector, at INFN - Laboratori Nazionali di Legnaro. Several new states with excitation
energy up to 8~MeV and spin up to $10^-$ have been discovered. These new states are discussed in terms of $J=3$ and $T=0$ neutron-proton hole pairs. Shell-model calculations 
in a large model space have shown a good agreement with the
experimental data for most of the energy levels.
The evolution of the structure of this nucleus is here studied 
as a function of excitation energy and angular momentum.
\end{abstract}
\maketitle

\section{\label{sec:intro}Introduction}

The region around the $N=20$ and $Z=20$ shell closures
has been subject of a number of experimental and theoretical investigations.
Single particle excitations, with configurations based on a spherical core corresponding to a shell closure, and collective excitations, in particular superdeformed rotational bands, are present in the \isotope{40}{Ca} \cite{PhysRevLett.87.222501} and \isotope{36}{Ar} \cite{PhysRevLett.85.2693,PhysRevC.63.061301,PhysRevC.65.034305} nuclei. These phenomena are the focus of a recently revived  experimental interest in this region \cite{springerlink:10.1140/epja/i2002-10125-6,kaisa}. These rotational structures are not only present in the even-even nuclei, but regular rotational bands 
of unnatural parity states have also been observed in odd-$A$ nuclei, for example in \isotope{43}{Ca}, \isotope{45}{Sc} and \isotope{45}{Ti} \cite{Styczen1976317,springerlink:10.1007/s100500050105,Bednarczyk} as well as odd-odd nuclei, for example \isotope{46}{V} \cite{PhysRevC.60.021303,OLeary199973}. 

While the primary goal of the present experiment was the performance evaluation of the AGATA detectors \cite{Soderstrom201196,alis,mythesis}, new results have been obtained for the reaction products, demonstrating the large sensitivity and efficiency of the AGATA spectrometer. In particular new results have been obtained for $^{40}$K using $\gamma\gamma$ coincidences 
employing just one AGATA triple cluster (ATC) detector.

The existing knowledge of high-spin states in \isotope{40}{K} is limited to the yrast band with $J^{\pi} \leq 9^{+}$ and two negative-parity states with $J^{\pi}=(8^-,10^-)$ and $J^{\pi}=(9^-,11^-)$, respectively. These high-spin states were studied using fusion-evaporation reactions, in particular: \reaction{\isotope{37}{Cl}}{\alpha}{\textrm{n}}{\isotope{40}{K}} \cite{cl1,cl2}, \reaction{\isotope{38}{Ar}}{\alpha}{\textrm{d}}{\isotope{40}{K}} \cite{DelVecchio1976220}, \reaction{\isotope{26}{Mg}}{\isotope{16}{O}}{\textrm{np}}{\isotope{40}{K}} \cite{k40_ppstates} and \reaction{\isotope{27}{Al}}{\isotope{19}{F}}{\alpha\textrm{np}}{\isotope{40}{K}} \cite{k40_1}. The positive-parity yrast states with $6^{+} \leq J^{\pi} \leq 9^{+}$ are well described in the shell
model as two particle and two hole states, with mainly a $(\mathrm{d}_{3/2}^{-2}\mathrm{f}_{7/2}^{2})$ configuration \cite{k40_ppstates}. Using the 
weak-coupling ansatz between particles and holes \cite{weak}
these states have been interpreted as part of a  $5^{+} \leq J^{\pi} \leq 9^{+}$ multiplet obtained by coupling a proton-hole pair $\mathrm{d}_{3/2}(J=2)$ to the $7^{+}$ state in \isotope{42}{Sc} \cite{k40_1}.
Furthermore, the same calculations predict a close lying  $3^{+} \leq J^{\pi} \leq 9^{+}$ multiplet from a
$^{42}$Ca($6^{+}$) core that couples to a neutron-proton hole-pair in the $\mathrm{d}_{3/2}$ orbital with $T=0$ and $J=3$ \cite{k40_1}.

The importance of isoscalar spin-aligned
neutron-proton pairs
for $N \sim Z$ nuclei has been the focus of much recent experimental and theoretical work at the $N=50$ and $Z=50$ shell
closures \cite{bosses_nature,isacker,PhysRevLett.107.172502,PhysRevC.84.021301,Xu2011,PhysRevC.85.034335}. In the $N=20$ and $Z=20$ region these pairs have also been discussed for Sc isotopes and related  
to the softening of the restoring force of dipole vibrations \cite{sc_isotopes}.

\section{Experiment and data processing}

The \reaction{\isotope{12}{C}}{\isotope{30}{Si}}{\mathrm{np}}{\isotope{40}{K}} reaction has been used during the commissioning phase of the AGATA $\gamma$-ray spectrometer
at INFN - Laboratori Nazionali di Legnaro (LNL) in Italy \cite{Soderstrom201196,Gadea201188}. A 64~MeV $^{30}$Si beam from the Tandem accelerator at INFN-LNL was used to
bombard a $200$~$\mu$g/cm$^2$ thick $^{12}$C target, producing $^{40}$K via the
fusion-evaporation reaction.  The $\gamma$ radiation was
detected by the first ATC detector
\cite{atc,agata}.
As the primary goal of this experiment was the performance evaluation of the AGATA detectors \cite{Soderstrom201196},
data were collected at two
distances between the front face of the ATC detector and
the target: a close setup with a distance of about 55~mm
and a far setup with a distance of about 235~mm.
The position of the detector with respect to the direction of the beam in the horizontal plane was 
$\theta_{\mathrm{beam}}=75.1$~degrees.
The beam intensities were about 0.3~pnA and 3~pnA for
the measurements at the close ($6.5{\cdot}10^8$ events) and the far ($2.8{\cdot}10^8$ events) distances, respectively. 

\begin{figure*}
 \centering
 \epsfig{file=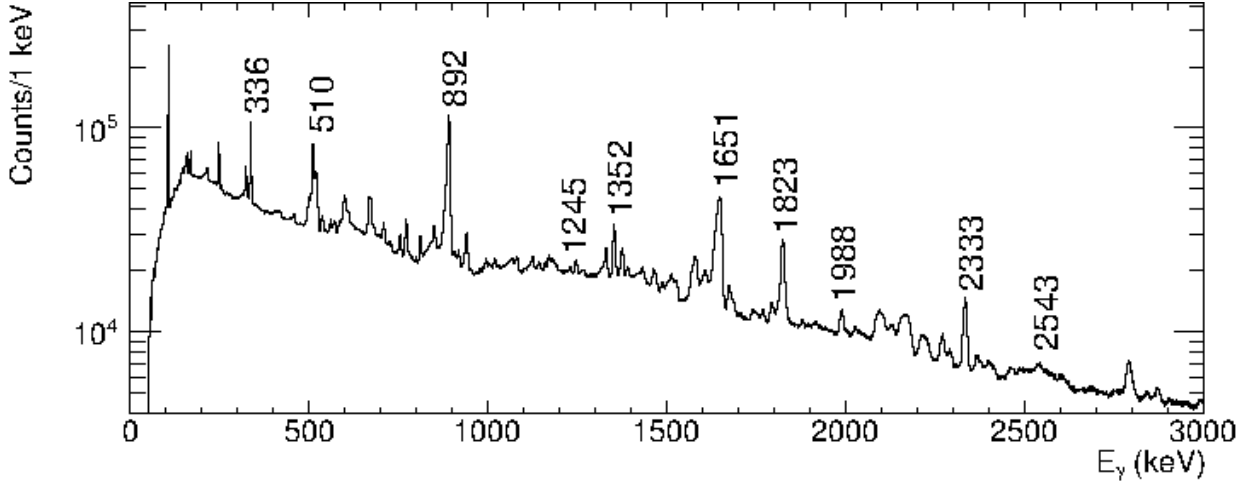,width=\textwidth}
 \caption{Total projection of the $\gamma\gamma$-coincidence spectrum. Known yrast transitions in  \isotope{40}{K} are labeled.}
 \label{fig:total_proj}
\end{figure*}

Doppler correction was carried out using the first interaction
point in the ATC as provided by the tracking algorithm, assuming an average recoil velocity of $v/c = 4.8$~\%. For further details about the experimental setup and the data processing, 
see \tref\cite{Soderstrom201196}.
After pulse-shape analysis and tracking, the reconstructed $\gamma$ rays were sorted into a $\gamma\gamma$-coincidence matrix. Using this procedure it was possible to use events where several $\gamma$ rays were detected in the same crystal, or scattered between crystals, in the analysis. The total projection of the resulting $\gamma\gamma$-coincidence spectrum is shown in \tfig\ref{fig:total_proj}.
It is evident from this spectrum that \isotope{40}{K},$\mathrm{np}$ is the
dominant reaction channel in the employed reaction with $\sim57$~\% of the $\gamma\gamma$-coincidence events. The main competing
channels are: \isotope{38}{Ar},$\alpha$ (16~\%); \isotope{40}{Ca},$2\mathrm{n}$; \isotope{40}{Ar},$2\mathrm{p}$; \isotope{37}{Cl},$\alpha\mathrm{p}$ (5~\% each); \isotope{41}{Ca},$\mathrm{n}$; \isotope{41}{K},$\mathrm{p}$; \isotope{37}{Ar},$\alpha\mathrm{n}$ (3.5~\% each) and \isotope{34}{S},$2\alpha$ (1.5~\%).

In addition to the already mentioned measurements at two distances, in order to search for short-lived isomers, a third measurements was performed with a 13~mg/cm$^2$
thick \isotope{58}{Ni} stopper foil placed $2.1$~mm downstream from the $200$~$\mu$g/cm$^2$ \isotope{12}{C} target.
During this measurement the ATC detector was placed at a distance of 
about 15 cm from the target and a total of $1.1{\cdot}10^7$ events were collected.

\section{Analysis}

\begin{figure*}
 \centering
\epsfig{file=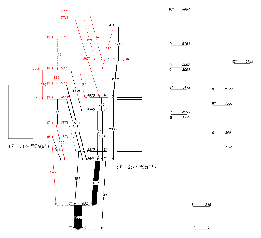,width=\textwidth}
 \caption{(Color online) Level scheme of \isotope{40}{K} as obtained in this work (left) and level scheme obtained from shell-model calculations (right). New information from this work is indicated with red color. The width of the arrows corresponds to the intensity of the $\gamma$ rays. For the positive-parity states, the assigned configurations in the weak-coupling basis are also shown.}
 \label{fig:level}
\end{figure*}

Several new $\gamma$ rays were assigned to \isotope{40}{K} based on the $\gamma\gamma$ coincidence relationships observed in this experiment. In particular, the following $\gamma$ rays, that were reported but not placed in the
level scheme of \tref\cite{k40_ppstates}, have been observed also in the present work: 811, 917, 939, 1526, 2269 and 2791~keV.  All $\gamma$ rays observed in the present work are shown in \tfig\ref{fig:level}  and summarized in  \ttab\ref{tab:gammas}. Three $\gamma$ rays that could not be placed in the level scheme were observed in this experiment at 323, 917 and 1792~keV.

\begin{table*}
\caption{Initial level energy, $E_{\mathrm{i}}$, and spin, $J^{\pi}_{\mathrm{i}}$, of the levels obtained in this work. For each $\gamma$ ray the energy $E_{\gamma}$, relative intensity $Br_{\gamma}$, singles intensity $I_{\gamma}$, final level energy $E_{\mathrm{f}}$, and final level spin $J^{\pi}_{\mathrm{f}}$, are listed. A systematic uncertainty of 0.2~keV has been added to the statistical uncertainty in $E_{\gamma}$. Similarly, a systematic uncertainty of 17~\% was added to $I_{\gamma}$. States and $\gamma$ rays labeled with $^\ast$ are observed for the first time in this work, while $\gamma$ rays labeled with $^{(\ast)}$ were reported in \tref\cite{k40_ppstates} but not placed in a level scheme in that work. Intensities labeled with $^{1}$ could not be measured in this work due to strong coincidences with \isotope{38}{Ar} transitions and have instead been obtained from \tref\cite{Cameron2004293}.\label{tab:gammas}}
 \begin{tabular*}{\textwidth}{@{\extracolsep{\fill}}lcccccc}
\hline
$E_{\mathrm{i}}$ & $J^{\pi}_{\mathrm{i}}$ & $E_{\gamma}$ & $Br_{\gamma}$ & $I_{\gamma}$ & $E_{\mathrm{f}}$ & $J^{\pi}_{\mathrm{f}}$\\
(keV) &  & (keV) &  &  & (keV) & \\
\hline
  $\val{891.75}{32}$  & ${5^{-}}$          & $\val{891.74}{22}$              & $\val{100}{18}$      & $\val{100}{18}$     & 0    & ${4^{-}}$      \\
  $\val{2543.2}{4}$ & ${7^{+}}$          & $\val{1651.34}{24}$             & $\val{100}{23}^1$    & $\val{61}{11}$      & 892  & ${5^{-}}$      \\
                      &                    & $\val{2543.2}{4}$               & $\val{12.6}{5}^1$  & $\val{4.5}{5}$   & 0    & ${4^{-}}$      \\
  $\val{2879.5}{4}$ & ${6^{+}}$          & $\val{336.25}{20}$              & $\val{100}{26}$      & $\val{7.2}{13}$  & 2543 & ${7^{+}}$      \\
                      &                    & $\val{1988.07}{35}$             & $\val{46}{12}$       & $\val{3.3}{6}$   & 892  & ${5^{-}}$      \\
  $\val{3354.0}{5}^{\ast}$ & $(6^{+})$   & $\val{810.79}{24}^{(\ast)}$     & $\val{100}{26}$      & $\val{1.75}{32}$ & 2543 & ${7^{+}}$      \\
                      &                    & $\val{2461.3}{11}^{\ast}$        & $\val{53}{14}$       & $\val{0.53}{17}$   & 892  & ${5^{-}}$      \\
  $\val{3872.7}{5}^{\ast}$ & $(7^{+})$   & $\val{518.97}{26}^{\ast}$         & $\val{6.9}{21}$     & $\val{0.23}{6}$   & 3354 & ${(6^{+})}$      \\
                      &                    & $\val{993.1}{4}^{\ast}$         & $\val{9.5}{25}$     & $\val{0.31}{6}$   & 2880 & ${6^{+}}$      \\
                      &                    & $\val{1329.00}{26}^{\ast}$      & $\val{100}{26}$      & $\val{3.3}{6}$   & 2543 & ${7^{+}}$      \\
  $\val{4366.1}{5}$ & ${(8^{+})}$        & $\val{1486.90}{34}$             & $\val{19}{6}^1$      & $\val{0.91}{17}$   & 2880 & ${6^{+}}$      \\
                      &                    & $\val{1822.83}{21}$             & $\val{100}{6}^1$     & $\val{13.3}{24}$  & 2543 & ${7^{+}}$      \\
  $\val{4812.4}{5}^{\ast}$ & $(8^{+})$   & $\val{939.28}{23}^{(\ast)}$     & $\val{100}{26}$      & $\val{5.7}{10}$   & 3873 & $(7^{+})$        \\
                      &                    & $\val{2269.0}{5}^{(\ast)}$      & $\val{73}{19}$       & $\val{4.1}{8}$   & 2543 & ${7^{+}}$      \\
  $\val{4876.0}{5}$ & ${9^{+}}$          & $\val{509.90}{20}$              & $\val{37}{10}$       & $\val{3.8}{7}$   & 4366 & ${(8^{+})}$    \\
                      &                    & $\val{2332.89}{22}$             & $\val{100}{26}$      & $\val{10.2}{19}$  & 2543 & ${7^{+}}$      \\
  $\val{5333.2}{5}^{\ast}$ & $(9^{+})$   & $\val{2790.53}{29}^{\ast}$        & $\val{100}{26}$      & $\val{5.4}{10}$   & 2543 & ${7^{+}}$      \\
  $\val{5892.2}{5}^{\ast}$ & $(9^{-})$   & $\val{559.28}{22}^{\ast}$       & $\val{62}{16}$       & $\val{0.57}{10}$ & 5333 & $(9^{+})$        \\
                      &                    & $\val{1016.6}{4}^{\ast}$        & $\val{48}{13}$       & $\val{0.44}{8}$   & 4876 & ${9^{+}}$      \\
                      &                    & $\val{1079.1}{5}^{\ast}$        & $\val{75}{20}$      & $\val{0.69}{13}$ & 4812 & $(8^{+})$        \\
                      &                    & $\val{1525.85}{27}^{(\ast)}$        & $\val{100}{27}$      & $\val{0.91}{18}$ & 4366 & ${(8^{+})}$         \\
\hline
 \end{tabular*} 
\end{table*} 
\begin{table*}
\caption{Continuation of \ttab\ref{tab:gammas}.\label{tab:gammas2}}
 \begin{tabular*}{\textwidth}{@{\extracolsep{\fill}}lcccccc}
\hline
$E_{\mathrm{i}}$ & $J^{\pi}_{\mathrm{i}}$ & $E_{\gamma}$ & $Br_{\gamma}$ & $I_{\gamma}$ & $E_{\mathrm{f}}$ & $J^{\pi}_{\mathrm{f}}$\\
(keV) &  & (keV) &  &  & (keV) & \\
\hline
  $\val{6227.5}{5}$ & $(10)^{-}$         & $\val{1351.70}{21}$             & $\val{100}{26}$      & $\val{7.3}{13}$  & 4876 & ${9^{+}}$      \\
  $\val{7033.4}{7}^{\ast}$ & $(9^{-})$   & $\val{1142.3}{5}^{\ast}$        & $\val{59}{16}$       & $\val{0.59}{11}$   & 5892 & $(9^{-})$        \\
                      &                    & $\val{2219.7}{5}^{\ast}$        & $\val{100}{26}$      & $\val{1.00}{18}$   & 4812 & $(8^{+})$        \\
  $\val{7472.3}{7}$ & ${(9^{-},11^{-})}$ & $\val{1245.10}{31}$             & $\val{21}{5}$        & $\val{1.12}{21}$   & 6228 & $(10)^{-}$ \\
                      &                    & $\val{1579.3}{5}^{\ast}$        & $\val{100}{26}$      & $\val{0.41}{9}$   & 5892 & $(9^{-})$        \\
  $\val{7748.5}{7}^{\ast}$ & $(9^{-},10^{-})$& $\val{1520.88}{30}^{\ast}$  & $\val{30}{9}$       & $\val{0.42}{9}$   & 6228 & $(10)^{-}$ \\
                      &                    & $\val{2872.9}{9}^{\ast}$        & $\val{100}{26}$      & $\val{1.42}{26}$   & 4876 & ${9^{+}}$      \\
  $\val{7994.6}{9}^{\ast}$ & $(9^{-}-12^{-})$& $\val{1767.1}{5}^{\ast}$    & $\val{100}{26}$      & $\val{0.60}{11}$ & 6228 & $(10)^{-}$ \\
\hline
 \end{tabular*} 
\end{table*} 
Since, as mentioned before, the main objective of the experiment was the detector performance, the particularities of the setup prevented to reproduce accurately the detector-target distances with calibration sources, consequently the direct measurement of efficiency calibration data was not possible. Instead, it was necessary to resort to Monte-Carlo simulations, performed with the \geant\
library \cite{Farnea2010331,Soderstrom201196,2003NIMPA.506..250G}, checked with efficiency measurements at standard distances and finally verified by comparing the obtained intensity ratios to previously measurements. This gives an estimated systematic uncertainty of the measured intensities of about $17$\%.

\begin{figure}
 \centering
 \epsfig{file=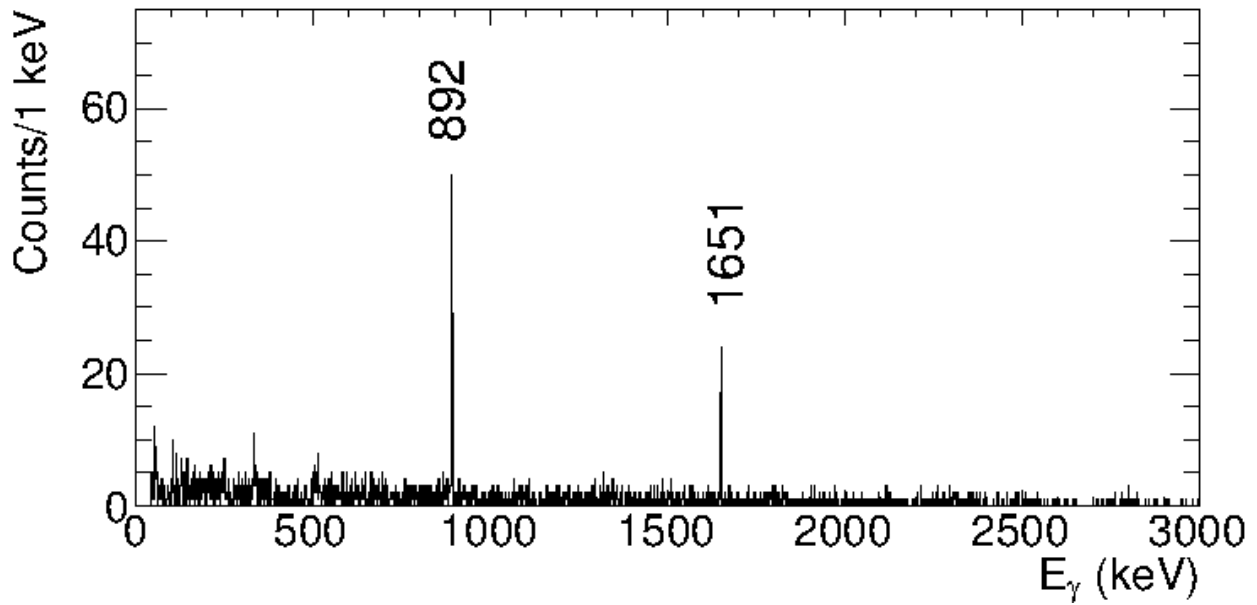,width=\columnwidth}
 \caption{Summed $\gamma\gamma$ coincidence spectrum measured with a 
  13~mg/cm$^2$ thick stopper foil of \isotope{58}{Ni} placed at a
  distance of $2.1$~mm from the target, without applying any Doppler correction. The spectrum is a sum of spectra gated on the 892~keV and the 1651~keV transitions.}
 \label{fig:stopped}
\end{figure}

The data set obtained with the \isotope{58}{Ni} stopper foil was analyzed in order
to search for short-lived isomers. The time it took for the
evaporation residues to reach the stopper foil was about $0.15$~ns. Any
$\gamma$ ray which is associated with an effective lifetime longer than
about 0.1 ns should, therefore, have narrow peaks in a spectrum
created without applying any Doppler correction. \tfig\ref{fig:stopped} shows a
$\gamma$-ray spectrum without Doppler correction that was created as the
sum of gates on the 892~keV and 1651~keV $\gamma$ rays. These $\gamma$ rays
correspond to two strong low-lying transitions in \isotope{40}{K} depopulating the
known $7^+$ isomer at 2543~keV (see the level scheme in \tfig\ref{fig:level}) with a
half-life of 1.09~ns \cite{Cameron2004293}. Since no other statistically significant peaks are visible in the
spectrum shown in \tfig\ref{fig:stopped}, no states could be identified in \isotope{40}{K} with
lifetimes longer than about~0.1 ns and which feed the 2543~keV state,

\subsection{Coincidence analysis}

\begin{figure}
 \centering
 \epsfig{file=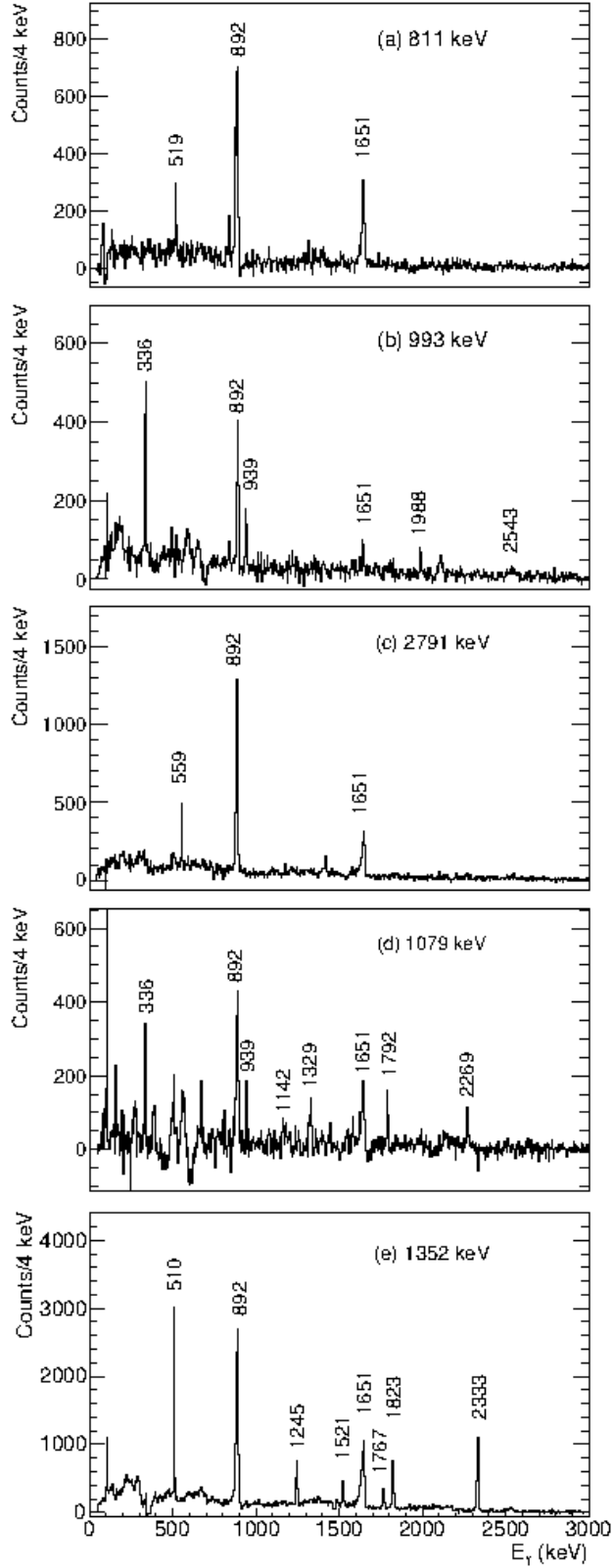,width=0.5\columnwidth}
 \caption{Background subtracted $\gamma\gamma$ spectra for some transitions in \isotope{40}{K}. The spectra have been obtained by requiring a 
coincidence relationship with the $\gamma$ transition reported above the spectrum.}
 \label{fig:gg_new}
\end{figure}

The level scheme in \tfig\ref{fig:level} was constructed using coincidence relationships between the observed $\gamma$ rays. Typical examples of $\gamma\gamma$-coincidence spectra from this experiment are shown in \tfig\ref{fig:gg_new}.

Our data confirm the previous placement and assignment of the excited 
levels up to spin $(10)^-$. In particular for all levels we could verify the 
coincidence relationships 
between transitions.

A new level is placed at an excitation energy of 3354~keV.
The line at 811~keV is in coincidence with the 1651-keV and the 892-keV $\gamma$ rays, see \tfig\ref{fig:gg_new}a, while the 2461-keV line is only in coincidence with 892~keV $\gamma$ ray and its energy corresponds within one keV to the sum of 811 and 1651~keV. In a similar way, a 993~keV $\gamma$ ray, see \tfig\ref{fig:gg_new}b, is observed in coincidence with the 336~keV, the 1988~keV, the 892~keV and the 939~keV transitions, while the $\gamma$ ray at 1329~keV is only in coincidence with the transitions at 1651~keV and 892~keV. Thus, a new level is placed at 3873~keV decaying to the $7^+_1$ state and the $6^+_1$ state.

Above the 3873~keV level, the 939~keV transition is in coincidence with the transitions at 993~keV and 1329~keV, implying that it decays from a new level at 4812 keV and feeds the 
3873 keV state. A $\gamma$ ray with an energy of
2269~keV is observed to be in coincidence with the 1651 and 892~keV 
$\gamma$ rays. It is placed between the new 4812~keV and the 
$7^+_1$ level at 2543~keV. Also, a $\gamma$ ray at 2269~keV is observed in coincidence with the 1651~keV $\gamma$ ray and the $\gamma$-ray of 892~keV, and thus this transition has the $6^+_1$ as final state, 
bypassing the level at 3873~keV.

A new level is placed at an excitation energy of 5333~keV on the basis of a 2791~keV transition that is observed depopulating this level, see \tfig\ref{fig:gg_new}c, and 
a $\gamma$ coincidence that has been established with the 1651 and 892~keV 
transitions, as expected. Moreover,
a 559~keV transition is observed in coincidence
with the three $\gamma$ rays at 2791~keV, 1651~keV and 892~keV. This transition was placed above the 5333~keV level, thus depopulating a 
newly placed level at an excitation energy of 5892~keV.
This choice is further supported by the observation of 
a $\gamma$ ray at 1079~keV in
coincidence with the 2269, 939, 993, 336,
1651 and 892 transitions, see \tfig\ref{fig:gg_new}d. Together with the 1526 and the 
 newly observed 
1017~keV transitions, the 1079~keV transition is assigned to depopulate the newly observed state at 5892~keV to the known $(8^+_1)$, $9^+_1$ and $(8^+_2)$ states at 4366, 4876 and 4812~keV respectively.

Finally a weak transition at 1142~keV is observed in coincidence with the transitions depopulating the two lowest lying excited states. Since this transition is weak and since the intensity of the decay, above the two states that appears in coincidence, is divided among several decay paths no statistically significant $\gamma$ coincidences with other transitions could be observed, except for the weak coincidence with the 1079~keV transition shown in \tfig\ref{fig:gg_new}d. Thus, together with the 2220~keV transition, this transition is tentatively assigned as the decay of a new 7033~keV level. The $\gamma$-ray of 2220~keV is as expected in coincidence with the 939~keV transition as well as with the 336~keV, 1651~keV and 892~keV $\gamma$ rays.

None of the $\gamma$ rays with energies 1521~keV and 1767~keV are in coincidence with the 1245~keV line, but all these three transitions are in coincidence with the yrast band, see \tfig\ref{fig:gg_new}e. The relative intensities of the 1245~keV, 1521~keV and 1767~keV transitions, with respect to the 1352~keV transition, are $\sim0.15$, $\sim0.058$ and $\sim0.082$, respectively. These small differences in the relative intensities suggests that all three of these transitions should decay to the state at 6228~keV. If the 1767~keV transition were to feed the 7748~keV level the probability to populate such a high-energy state, and hence the intensity of the 1767~keV transition, would be expected to drop dramatically. The entire yrast band up to the $9^{+}$ level is also in coincidence with the 2873~keV transition, corresponding within one keV to the sum of 
the energies of the transitions at 1352 and 1521~keV. This coincidence further strengthen the placement of the $\gamma$ ray at 1521~keV.

\begin{figure*}
 \centering
 \epsfig{file=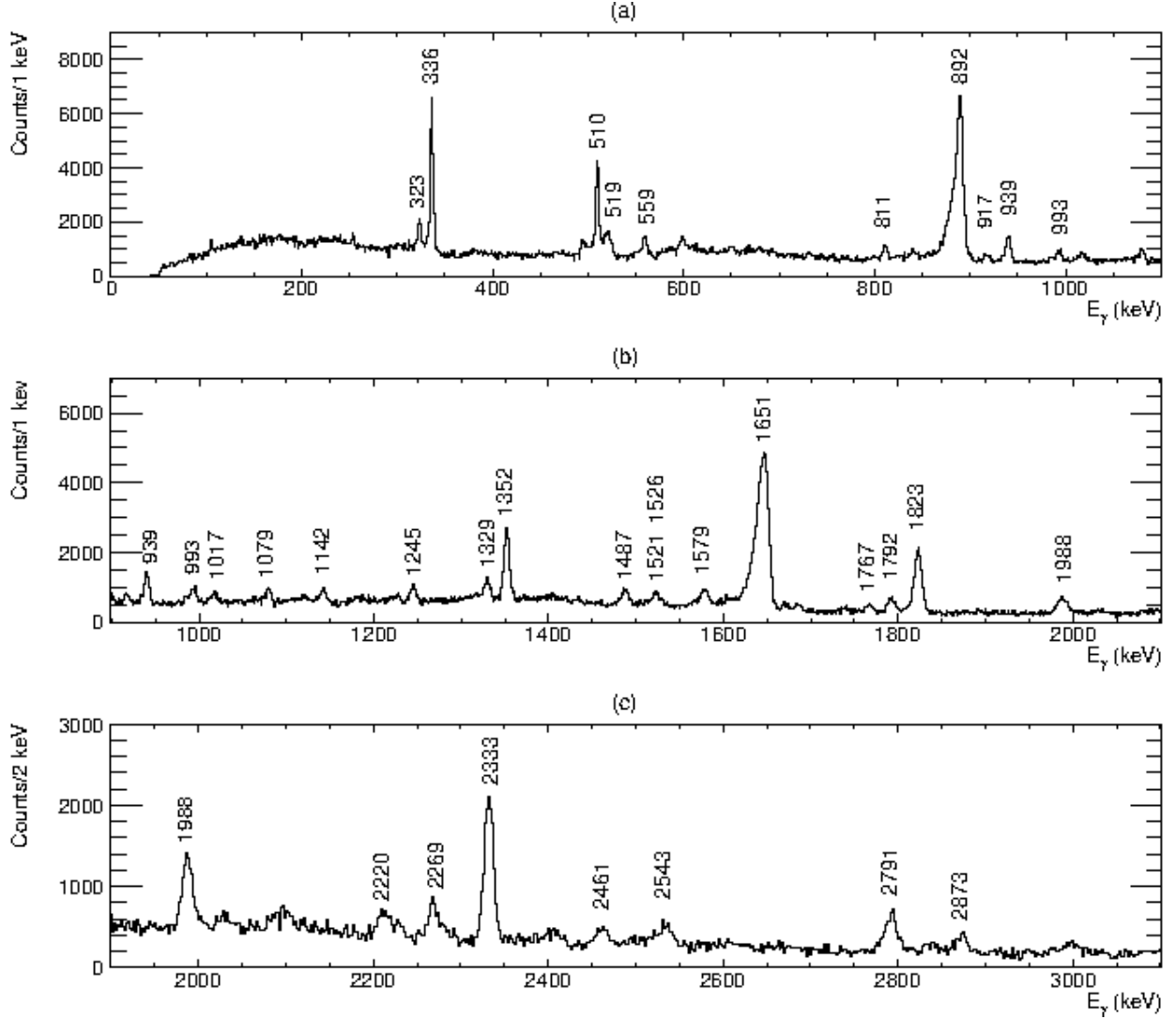,width=\textwidth}
 \caption{Summed $\gamma\gamma$-coincidence spectrum for all transitions listed in \ttab\ref{tab:gammas} in the energy ranges 0--1100~keV (a), 900--2100~keV (b) and 1900--3100~keV (c). The transitions at 1651 and 1823~keV  were excluded from the 
gating conditions because similar energies are present in the \isotope{38}{Ar} decay scheme. The low-energy tail of the 892 and
1651~keV peaks are due to the 1.09~ns isomer of the $7^+_1$
state.}
 \label{fig:known}
\end{figure*}

A summed coincidence spectrum for the observed $\gamma$ rays with an energy up to 3000~keV is shown in \tfig\ref{fig:known}. 

\subsection{$J^{\pi}$ assignment}

Tentative spin and parity assignments have been made of some of the new levels.
Unfortunately the limited angle subtended by the detector prevented the 
measurement of angular distributions
and thus the assignments had to rely on the comparison of the branching ratios to Weisskopf 
estimates.
The relative intensity of the transitions  
were estimated
according to \tref\cite{transdata}. In order to obtain an estimation of the intensity, $I_{\mathrm{W}}$, of the $\gamma$ ray for a given multipolarity, the decay widths have to be multiplied with the strength, $S$, of the transition.

The transition strengths, $S$, are not known explicitly for each level, but under the assumption that they are similar for similar types of transition in a given nucleus they can be estimated from previously known intensities in the given nucleus. Using the relative intensities and mixing ratios reported in \tref\cite{Cameron2004293}, the strength parameters have been adjusted to reproduce known intensities. Fixing $S(\mathrm{E}2) = 10.0$ as reference value, the best values for the transition strengths obtained were $S(\mathrm{E}1) = 3.65\cdot10^{-5}$ and  $S(\mathrm{M}1) = 5.42\cdot10^{-2}$. These values are consistent with the strength distributions of $\gamma$-ray transitions for $A=5$--$40$ shown in \tfig2 of \tref\cite{transdata}. 

With these values it was possible to reproduce known intensities within a factor of three in most cases, and within a factor of ten
in the worst case. When the wrong parity was used for known states, the intensities differed three to five orders of magnitude.

A tentative assignment
of spin and parity was possible for the lowest lying
levels discovered in this work assuming a
mixing ratio of $\delta=0$ for the transitions. 
These assignments are
shown in \ttab\ref{tab:spinparity}.
\begin{table}
\caption{Tentative spin and parity assignments, $J^{\pi}_{\mathrm{i}}$, for the lowest lying energy levels observed in this work. Only the observed $\gamma$ rays have their measured intensity ($Br_{\gamma}$) listed while both observed and possible but unobserved $\gamma$ rays have their energies ($E_{\gamma}$) and calculated intensities ($Br_{\gamma}^{\mathrm{W}}$) listed. The multipolarity (Mult.) used for the calculation and the spin-parity ($J^{\pi}_{\mathrm{f}}$) are listed for both observed and unobserved transitions. This table shows only the new states, the previously known states are not included. \label{tab:spinparity}}
 \begin{tabular*}{\columnwidth}{@{\extracolsep{\fill}}lccccr}
\hline
$J^{\pi}_{\mathrm{i}}$ & $E_{\gamma}$ & $Br_{\gamma}$ & $Br_{\gamma}^{\mathrm{W}}$ & Mult. & $J^{\pi}_{\mathrm{f}}$\\
 & (keV) & & & & \\
\hline
$(6^+)$ & 474 &  & 20 & M1 & $6^+$\\
 & 811 & $\val{100}{26}$ & 100 & M1 & $7^+$\\
 & 2461 & $\val{53}{14}$ & 71 & E1 & $5^-$\\
$(7^+)$ & 519 & $\val{6.9}{21}$ & 5.9 & M1 & $(6^+)$\\
 & 993 & $\val{9.5}{25}$ & 42 & M1 & $6^+$\\
 & 1329 & $\val{100}{26}$  & 100 & M1 & $7^+$\\
$(8^+)$ & 446 &  & 0.8 & M1 & $(8^+)$\\
 & 939 & $\val{100}{26}$ & 7.1 & M1 & $(7^+)$\\
 & 1458 &  & 3.3 & E2 & $(6^+)$\\
 & 1933 &  & 14 & E2 & $6^+$\\
 & 2269 & $\val{73}{19}$ & 100 & M1 & $7^+$\\
$(9^+)$ & 457 &  & 1.0 & M1 & $9^+$\\
 & 521 &  & 1.4 & M1 & $(8^+)$\\
 & 967 &  & 9.1 & M1 & $(8^+)$\\
 \hline
 \end{tabular*} 
\end{table} 
\begin{table}
\caption{Continuation of \ttab\ref{tab:spinparity}\label{tab:spinparity2}}
 \begin{tabular*}{\columnwidth}{@{\extracolsep{\fill}}lccccr}
\hline
$J^{\pi}_{\mathrm{i}}$ & $E_{\gamma}$ & $Br_{\gamma}$ & $Br_{\gamma}^{\mathrm{W}}$ & Mult. & $J^{\pi}_{\mathrm{f}}$\\
 & (keV) & & & & \\
\hline
 & 1461 &  & 3.9 & E2 & $(7^+)$\\
 & 2791 & $\val{100}{26}$ & 100 & E2 & $7^+$\\
$(9^-)$ & 559 & $\val{62}{16}$ &  4.9 &  E1 &  $(9^+)$\\
 &  1017 &  $\val{48}{13}$&  30 &  E1 &  $9^+$\\
 &  1079 &  $\val{75}{20}$  &  35 &  E1 &  $(8^+)$\\
 &  1526 &  $\val{100}{27}$ &  100 &  E1 &  $(8^+)$\\
$(10)^{-}$ & 894 &  &  29 &  E1 &  $(9^+)$\\
 &  1352 &  $\val{100}{26}$ &  100 &  E1 &  $9^+$\\
$(9^-)$ & 806 &  & 35 & M1 & $(10)^{-}$\\
 & 1142 & $\val{59}{16}$ & 100 & M1 & $(9^-)$\\
 & 1700 &  & 8.4 & E1 & $(9^+)$\\
 & 2157 &  & 17 & E1 & $9^+$\\
 & 2220 & $\val{100}{26}$ & 19 & E1 & $(8^+)$\\
 & 2667 &  & 33 & E1 & $(8^+)$\\
 \hline
 \end{tabular*} 
\end{table} 

For the states at 7472~keV, 7748~keV and 7995~keV it was not possible to unambiguously restrict the spin using this method.

\section{Discussion}

As mentioned in \tsec1, \tref\cite{k40_1} interprets the yrast positive-parity 
states with $6^{+} \leq J^{\pi} \leq 9^{+}$ as arising from a proton-hole pair $(J=2)$ in the $\mathrm{d}_{3/2}$ orbital coupled to the $7^{+}$ state in $^{42}$Sc. In the same reference, the weak coupling calculations also predict a positive-parity multiplet with $3^{+} \leq J^{\pi} \leq 9^{+}$ due to a $^{42}$Ca($6^{+}$) core that couples to a neutron-proton hole-pair in the $\mathrm{d}_{3/2}$ orbital with $T=0$ and $J=3$. The highest spin state in this multiplet, $9^{+}$, is predicted at an excitation energy of 5.23~MeV, which is just slightly lower than the $(9^{+})$
state at 5333~keV, tentatively identified in the present work.
For the high-energy negative-parity states, several different configurations, lying close in energy, could be assigned. Below 8~MeV the following $\mathrm{d}_{3/2}^{-3}(T,J)$ multiplets with isospin $T$, spin $J$ and excitation
energy $E(J^{\pi}_{\mathrm{max}})$ of the highest spin state in the multiplet were predicted in \tref\cite{k40_1}:
\begin{itemize}
 \item $\mathrm{d}_{3/2}^{-3}(\frac{1}{2},\frac{3}{2})\otimes\isotope{43}{Sc}(\frac{17}{2}^-)$, $E(10^-) = 7.96$~MeV,
 \item $\mathrm{d}_{3/2}^{-3}(\frac{3}{2},\frac{3}{2})\otimes\isotope{43}{Ti}(\frac{19}{2}^-)$, $E(11^-) = 7.50$~MeV,
 \item $\mathrm{d}_{3/2}^{-3}(\frac{1}{2},\frac{3}{2})\otimes\isotope{43}{Ca}(\frac{15}{2}^-)$, $E(9^-) = 6.75$~MeV,
 \item $\mathrm{d}_{3/2}^{-3}(\frac{1}{2},\frac{3}{2})\otimes\isotope{43}{Sc}(\frac{19}{2}^-)$, $E(11^-) = 6.72$~MeV,
 \item $\mathrm{d}_{3/2}^{-3}(\frac{1}{2},\frac{3}{2})\otimes\isotope{43}{Sc}(\frac{15}{2}^-)$, $E(9^-) = 6.59$~MeV.
\end{itemize}
It is not possible to unambiguously assign the observed high-energy negative-parity states to any of these configurations, or combination of configurations, using the current data.

Shell-model calculations have been carried out using the code ANTOINE \cite{poves2,poves3} in the valence space comprising the
orbits $1\mathrm{s}_{1/2}$, $0\mathrm{d}_{3/2}$, $0\mathrm{f}_{7/2}$, $1\mathrm{p}_{3/2}$, $1\mathrm{p}_{1/2}$ and $0\mathrm{f}_{5/2}$ for neutrons and protons. The
valence space and the effective interactions have been demonstrated successful in the description of \isotope{40}{Ca} in \tref\cite{poves1}.
Up to six particles are allowed to move from the sd  to the pf shell.
The results of these calculations for the yrast and yrare states with $6 \leq J \leq 10$ are shown in \ttab\ref{tab:smcalc} and \ref{tab:smcalcneg} and in \tfig\ref{fig:level}.
\begin{table*}
\caption{Experimental energies ($E_{\mathrm{exp}}$), calculated energies ($E_{\mathrm{th}}$) and amount of contribution from different configurations of the positive-parity energy-levels according to shell-model calculations. The energy levels have been sorted into two groups depending on whether the dominant configuration is $(\pi\mathrm{d}_{3/2}^{-2}\mathrm{f}_{7/2}^{1})\otimes(\nu\mathrm{f}_{7/2}^{1})$ (top) or $(\pi\mathrm{d}_{3/2}^{-1})\otimes(\nu\mathrm{d}_{3/2}^{-1}\mathrm{f}_{7/2}^{2})$ (bottom). \label{tab:smcalc}}
 \begin{tabular*}{\textwidth}{@{\extracolsep{\fill}}lcccccc}
\hline
$J^{\pi}_{n}$ & $E_{\mathrm{exp}}$ & $E_{\mathrm{th}}$ & $(\pi\mathrm{d}_{3/2}^{-2}\mathrm{f}_{7/2}^{1})\otimes(\nu\mathrm{f}_{7/2}^{1})$ & $(\pi\mathrm{d}_{3/2}^{-1})\otimes(\nu\mathrm{d}_{3/2}^{-1}\mathrm{f}_{7/2}^{2})$ & $(\pi\mathrm{d}_{3/2}^{-2}\mathrm{f}_{7/2}^{1})\otimes(\nu\mathrm{d}_{3/2}^{-2}\mathrm{f}_{7/2}^{3})$ & $(\pi\mathrm{d}_{3/2}^{-1})\otimes(\nu\mathrm{d}_{3/2}^{-1}\mathrm{f}_{7/2}^{1}\mathrm{p}_{3/2}^{1})$\\
 & (keV) & (keV) & (\%) & (\%) & (\%) & (\%)\\
\hline
 $\phantom{(}6^{+}_{1}\phantom{)}$ & 2880 & 3500 & 17 & 18 & 18 & -\\
 $\phantom{(}7^{+}_{1}\phantom{)}$ & 2543 & 2976 & 58 & - & - & -\\
 $(8^{+}_{2})$ & 4812 & 5172 & 58 & - & - & -\\
 $\phantom{(}9^{+}_{1}\phantom{)}$ & 4876 & 5121 & 41 & 30 & - & -\\
 \hline
 $(6^{+}_{2})$ & 3354 & 4222 & 10 & 34 & - & 13\\
 $(7^{+}_{2})$ & 3873 & 4255 & - & 53 & - & -\\
 $(8^{+}_{1})$ & 4366 & 4586 & 10 & 58 & - & -\\
 $(9^{+}_{2})$ & 5333 & 5992 & 13 & 31 & - & -\\
\hline
 \end{tabular*} 
\end{table*} 
\begin{table*}
\caption{Experimental energies ($E_{\mathrm{exp}}$), calculated energies ($E_{\mathrm{th}}$) and amount of contribution from different configurations of the negative-parity energy-levels according to shell-model calculations. \label{tab:smcalcneg}}
 \begin{tabular*}{\textwidth}{@{\extracolsep{\fill}}lccccc}
\hline
$J^{\pi}_{n}$ & $E_{\mathrm{exp}}$ & $E_{\mathrm{th}}$ & $(\pi\mathrm{d}_{3/2}^{-1})\otimes(\nu\mathrm{f}_{7/2}^{1})$ & $(\pi\mathrm{d}_{3/2}^{-2}\mathrm{f}_{7/2}^{1})\otimes(\nu\mathrm{d}_{3/2}^{-1}\mathrm{f}_{7/2}^{2})$ & $(\pi\mathrm{d}_{3/2}^{-2}\mathrm{f}_{7/2}^{1})\otimes(\nu\mathrm{d}_{3/2}^{-1}\mathrm{f}_{7/2}^{1}\mathrm{p}_{3/2}^{1})$\\
 & (keV) & (keV) & (\%) & (\%) & (\%)\\
\hline
 $\phantom{(}4^{-}_{1}\phantom{)}$ & 0 & 0 & 70 & - & - \\
 $\phantom{(}5^{-}_{1}\phantom{)}$ & 892 & 846 & 69 & - & - \\
 $(9^{-}_{1})$ & 5892 & 6007 & - & 42 & 12 \\
 $\phantom{(}10^{-}_{1}\phantom{)}$ & 6228 & 6133 & - & 60 & - \\
 $(9^{-}_{2})$ & 7033 & 6790 & - & 41 & 12 \\
 $(10^{-}_{2})$ & - & 8057 & - & 30 & 21 \\
\hline
 \end{tabular*} 
\end{table*} 
The shell-model calculations reproduce the experimental energies well, with the exception of the $J^{\pi} = 6^{+}_{1,2},9^{+}_{2}$ states that have lower energies than the theoretical predictions. These states are also the states with the strongest predicted mixing between different configurations.

The shell-model calculations also reproduce the branching ratios well, except for the transitions to and from the $J^{\pi} = 6^{+}_{1,2},9^{+}_{2}$ states. As seen in \ttab\ref{tab:smcalc}, the interpretation of the yrast states dominantly belonging to a $(\pi\mathrm{d}_{3/2}^{-2}\mathrm{f}_{7/2}^{1})\otimes(\nu\mathrm{f}_{7/2}^{1})$ configuration corresponding to a $J=2$ multiplet coupled to the $7^{+}$ state in $^{42}$Sc in the weak coupling scheme, is supported by the shell-model calculations. The interpretation of the yrare states dominantly belonging to a $(\pi\mathrm{d}_{3/2}^{-1})\otimes(\nu\mathrm{d}_{3/2}
^{-1}\mathrm{f}_{7/2}^{2})$ configuration, corresponding to a $T=0$ and $J=3$ multiplet coupled to the $6^{+}$ state in $^{42}$Ca in the weak coupling basis, is supported by the calculations. Exceptions are the $8^{+}_{1,2}$ states, where the $8^{+}_{1}$ has a larger contribution of the $(\pi\mathrm{d}_{3/2}^{-1})\otimes(\nu\mathrm{d}_{3/2}
^{-1}\mathrm{f}_{7/2}^{2})$ configuration and the $8^{+}_{2}$ a larger contribution of the $(\pi\mathrm{d}_{3/2}^{-2}\mathrm{f}_{7/2}^{1})\otimes(\nu\mathrm{f}_{7/2}^{1})$ configuration. 
We can then associate the $7^{+}_{1}$, $8^{+}_{2}$ and $9^{+}_{1}$ states to the coupling of a pair of proton holes with $J=2$ to the $^{42}$Sc $7^{+}$ state, while the $6^{+}_{2}$, $7^{+}_{2}$, $8^{+}_{1}$ and $9^{+}_{2}$ states agree with the weak coupling of a proton-neutron aligned $(J=3)$ $T=0$ pair to the $6^{+}$ state in $^{42}$Ca.

The negative-parity states at high excitation energy are clearly dominated by the $(\pi\mathrm{d}_{3/2}^{-2}\mathrm{f}_{7/2}^{1})\otimes(\nu\mathrm{d}_{3/2}^{-1}\mathrm{f}_{7/2}^{2})$ configuration, with a small mixing of the $(\pi\mathrm{d}_{3/2}^{-2}\mathrm{f}_{7/2}^{1})\otimes(\nu\mathrm{d}_{3/2}^{-1}\mathrm{f}_{7/2}^{1}\mathrm{p}_{3/2}^{1})$ configuration. This is consistent with three holes outside a \isotope{43}{Sc} core in the weak-coupling picture. However, there are three different possible configurations of the \isotope{43}{Sc} core in this region, and it is not possible to discriminate between these using the current data.

\section{Summary}

High-spin states in \isotope{40}{K} have been studied via $\gamma$-ray spectroscopy using one AGATA triple cluster during the AGATA commissioning experiment at INFN-LNL. Several new states with excitation energy up to 8~MeV and spin $J\leq9$ have been discovered. Shell-model calculations in a large model space, which include orbitals of two main shells, reproduce well 
the experimental findings. The new states observed can be interpreted as the weak coupling of a proton-neutron aligned $(J=3)$ $T=0$ pair to the $6^{+}$ state in $^{42}$Ca.

\begin{acknowledgments}
We would like to thank the Swedish National Infrastructure for
Computing (SNIC) and the Uppsala Multidisciplinary Center for Advanced
Computational Science (UPPMAX) for the computer
resources used in parts of the analysis. This work was financed by EURONS AGATA (Contract
no. 506065-R113), the Swedish Research Council under contracts 621-2008-4163, 621-2011-4522, 822-2005-3332 and 821-2010-6024, the Japan Society for the Promotion of Science (JSPS) Kakenhi Grant No.~23$\cdot$01752, Ankara
University (BAP Project number 05B4240002), the Polish Ministry of Science and Higher Education 
under grants number DPN/N190/AGATA/2009 and N~N202~073935, the German BMBF under Grants 06K-167, 06KY205I and 06KY9136I, the Knut and Alice Wallenberg Foundation contract number 2005.0184 and STFC (UK).
A.~Gadea and E.~Farnea acknowledge the support of MICINN, Spain, and INFN, Italy through the AIC-D-2011-0746 bilateral action. A.~Gadea's activity has been partially supported by the Generalitat Valenciana, Spain, under grant PROMETEO/2010/101.
A.~Gadea, A.~Jungclaus, A.~Poves and B.~Quintana acknowledge support from the MICINN, Spain, under grant FPA2011-29854. A.~Pove is partially supported by the Comunidad de Madrid (Spain) (HEPHACOS S2009-ESP-1473).  G. Jaworski acknowledges the support of the
Center for Advanced Studies of Warsaw University of Technology. B.~Cederwall acknowledge the support of the G\"{o}ran Gustafsson Foundation.
\end{acknowledgments}


\end{document}